\documentstyle{EuroPhys}
\input epsf
\input EuroMacr
\begin{document}
%
%%%   The headers.
%
%%%   These three macros are to have correct headings in your paper.
%%%   You shall omit all the arguments in the two macros `\euro{}{}{}{}'
%%%   `\Date{}' and fill in `\shorttitle{}'. 
%%%   If there is more than one author in the 
%%%   \shorttitle macro, use the macro \etal after first author's name
%%%   to obtain the correct heading.
%
\euro{}{}{}{}
\Date{}
\shorttitle{M. I. ANTONOYIANNAKIS \etal MIE RESONANCES AND BONDING ETC.}
%
%%%  The title, the Author(s) and the affiliation(s)
%
%%%   The title is set in bold (initial word only is capitalized).
%%%   Mathematical expressions and formulas within the title shall be left
%%%   in light face. Initial(s) of the first name(s) are followed by the
%%%   author(s)'s last name(s). If the authors have different affiliations,
%%%   the name must be followed by one or more \inst{number} each referring
%%%   to one of the addresses to appear in the following macro \institute.
%%%   Other items like `Present address' or `email' may be added by putting
%%%   a `\footnote' after the last \inst{number}.
%%%   Begin each address with \inst{number}; the end of an address is \\;
%%%   \\ can also be used to break a line.
%
\title{Mie resonances and bonding in photonic crystals}
\author{M. I. Antonoyiannakis\footnote{BEIT SCIENTIFIC RESEARCH FELLOW}, J. B. Pendry}
\institute{
      Condensed Matter Theory Group, Blackett Laboratory, 
              Imperial College, London SW7 2BZ, UK}
%
%%%    The `\maketitle' macro needs the following macro:    \rec{}{}
%%%    to be left empty.
%
\rec{}{}
%
%%%   Physics Abstracts Classification.
%
%%%   There are two macros: the first one `\pacs{}' makes the PACS 
%%%   environment,the second one `\Pacs{}{}{}' can be used for each
%%%   classification you need.
%%%   To create the subject index of the volume it is important to divide
%%%   the classification numbers into the three different arguments like
%%%   in the following examples 
%
\maketitle
%
%%%   ! Don't forget this command to format the title page of your article!
%
%%%   The Abstract
%
\begin{abstract}
Isolated dielectric spheres support resonant electromagnetic (EM) modes which are analogous to electronic orbitals and, like their electronic counterparts, can form bonding or anti-bonding interactions between neighbouring spheres. By irradiating the system with light at the bonding frequency an attractive interaction is induced between the spheres. We suggest that by judicious selection of bonding states we can drive a system towards a desired structure, rather than rely on the structure dictated by gravitational or Van der Waals forces, the latter deriving from the zero point energy population of a state.
\end{abstract}
%
%
%%%   Main text
%
%%%   Sectioning
%
%%%   In EuroPhys there is only ``one'' level of sectioning `\section{}'.
%
% Main text begins here.
%
\section{Introduction}
The driving force behind our work is the understanding of EM forces in matter at the nanometre and micron scales. The biologically important Van der Waals force in solids belongs to this category. Developments in photonic band gap materials and colloidal crystals (as for example in \cite{PBG}) have made it necessary to study such systems from the point of view of stability and cohesion. Up to now efforts to fabricate photonic crystals with tailored optical properties in the visible range of the spectrum have stumbled on the experimental difficulty in assembling crystal layers into fully three-dimensional (3D) structures, although some promising methods have arisen \cite{3D_in_100nm}. We report here on an idea which may facilitate formation of such structures, as well as being of physical interest in its own right.

In this Letter we study the resonant EM modes in photonic crystals and their implications for EM forces. The origin of such modes lies in the EM field oscillations of an isolated sphere under the influence of an external plane-polarised radiation field ({\it Mie resonances}). Modes of similar nature (but corresponding to elastic waves in periodic composite materials) were identified by Kafesaki {\it et al} \cite{Kaf}. In the case of light in photonic crystals such resonant modes were first demonstrated (to our knowledge) by Zhang and Satpathy \cite{Zhang}, whereas Stefanou {\it et al} \cite{SKM} commented on their excitability by incident light. Recently Ohtaka and Tanabe \cite{Oht} studied them thoroughly, appropriately calling them {\it heavy photons} due to their low dispersion behaviour; we adopt their terminology. We show that such modes result in sharp features in the force spectrum, causing mutual attraction or repulsion between successive crystal layers. 

The basic results of Mie scattering theory \cite{Mie} will be needed so we present them here briefly. When plane-polarised light of frequency $\omega$ is incident on a sphere, a resonance situation is possible whenever $\omega$ matches the frequency of an EM eigenmode of the sphere. The eigenmodes are spherical waves labelled by three quantum numbers, just as in the case of electronic orbitals in atoms, and are generally electric or magnetic oscillating multipoles \footnote {Some differences exist in the nature and values of these EM quantum numbers with respect to the electronic case, but they are of no significance at this stage.}. For example the lowest in frequency excitable mode by a plane wave incident on a dielectric sphere is a magnetic dipole mode, labelled as $b_1^1$. Since these resonant modes are the EM equivalent of the atomic energy levels, it is only natural to expect that when two spheres supporting such modes come close together, hybridisation of the modes and bonding effects may occur.  

\section{Methods of calculation}
Our numerical method consists of discretising Maxwell's equations on a simple cubic (SC) mesh. The transfer matrix formalism is adopted in calculating the fields in real space. Our computer codes have been published and explained in detail elsewhere \cite{OPAL}. The new ingredient is the calculation of forces. Towards this end we adopt the {\it Maxwell Stress Tensor} (MST) methodology, whereby the time-averaged force on a body within a volume element is found by integrating the MST over the {\it closed} surface of the volume element:
 				\begin{equation}
 F_\alpha = \int T_{\alpha \beta} \,dS_\beta  \label{eq:MSTdef_force} \;\;\;\;\;\;\;\;
\alpha, \beta = \{x,y,z\} 
				\end{equation} 
where a sum over repeated indices is implied. The time-averaged stress tensor is given by
				\begin{equation} 
T_{\alpha \beta} = E_{\alpha} \,D_{\beta}^* + H_{\alpha} \,B_{\beta}^* -
\frac{1}{2} \,\delta_{\alpha \beta} \,( E_{\gamma} \,D_{\gamma}^* +
 H_{\gamma} \,B_{\gamma}^* )  \label{eq:MSTdef} \,.
				 \end{equation}
Therefore if we know the field around an object we can calculate the full EM force acting on it. We have implemented this algorithm in our computer codes. A complete description of this methodology is beyond the scope of this Letter but will be given elsewhere\footnote{Work in preparation}.

\section{Results and discussion}
Since we wish to study EM effects on dielectric crystals, we chose as a model system a crystal of Gallium Phosphide (GaP) spheres in a SC lattice. GaP has an essentially real, constant dielectric function (modelled here as $\epsilon=8.9$) over the frequency range of relevance to this work (0.07-0.9 eV). We have checked that the absorption of GaP is too small to have any sizeable effect on our force calculations. We have chosen a radius $r = 365$ nm and a lattice spacing $l = 900$ nm, so the spherical surfaces are separated by $D=170$ nm. For simplicity we consider the reference medium in which the spheres are embedded to be air, although, for reasons of experimental feasibility, it is really a liquid medium we have in mind. Such a medium does not alter qualitatively the effects we describe provided its absorption is low enough.
%(for example water will do).
Most of our results are for a discretisation mesh $10\times10\times10$.

For the EM properties of this crystal we expect the following behaviour: (a) propagation of light in an ``effective medium" at large wavelengths $\lambda \gg l$ where light senses the crystal as an homogeneous dielectric medium; (b) appearance of forbidden ranges of frequencies as $\lambda$ %
% fig1
\begin{figure}
%%%%%%%%\vskip -0.4in
\epsfxsize=2.5in
\epsfysize=4.5in
%\epsffile{plots/fig1/pPBSTRANS.ps}
\epsffile{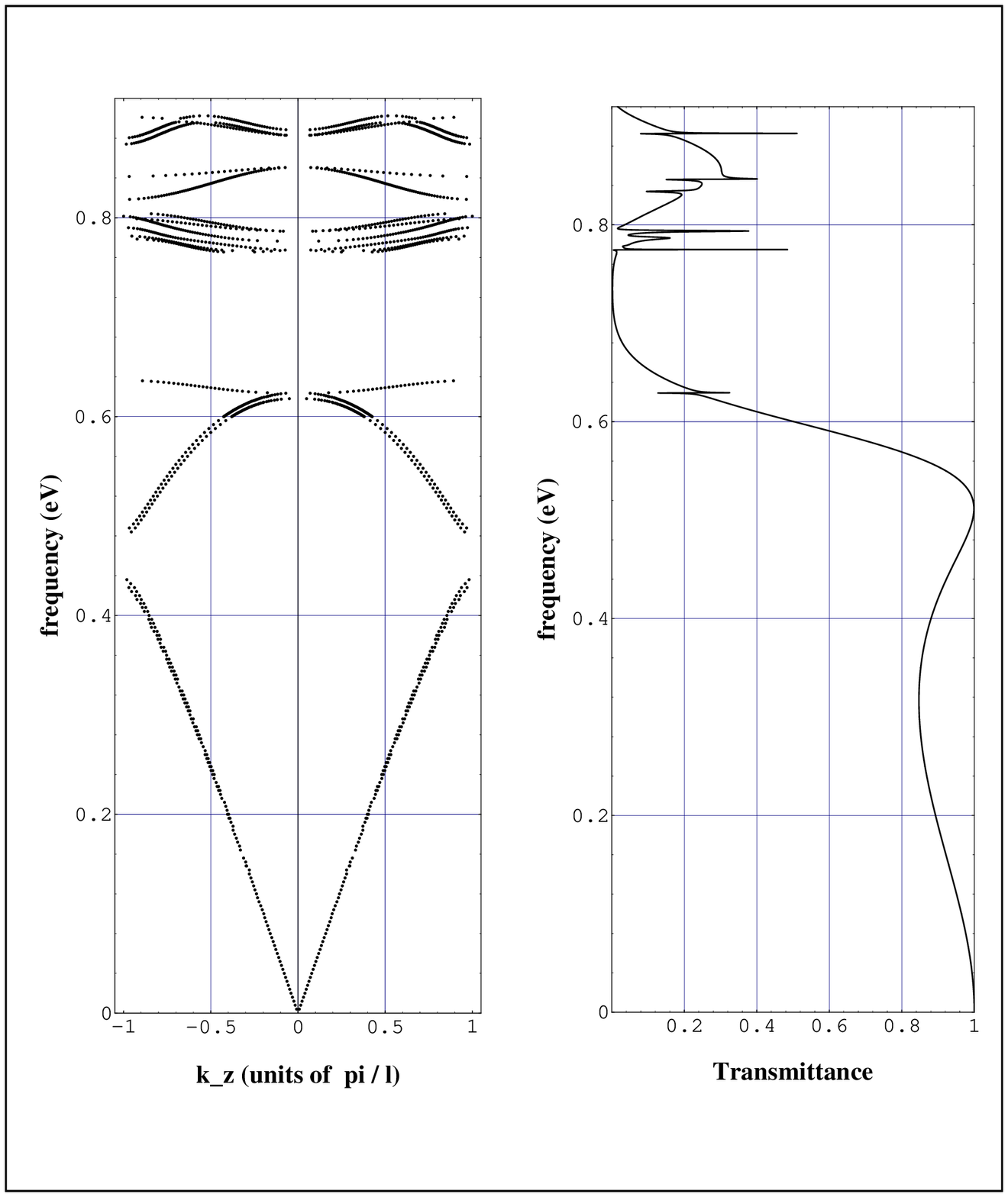}
\vskip -0.4in
\caption{Photonic band structure along the $\Gamma X$ direction for a SC lattice and transmittance for a crystal one layer thick.}
\label{fig1}
%\end{figure}
% 
%fig2a
%
%\begin{figure}
\epsfxsize=1.5in
%\epsffile{plots/fig2/a/y7.H.s.ps}
%\special{psfile=plots/fig2/a/y8.H.s.ps hoffset=135 voffset=150 hscale=35 vscale=35}
\includegraphics{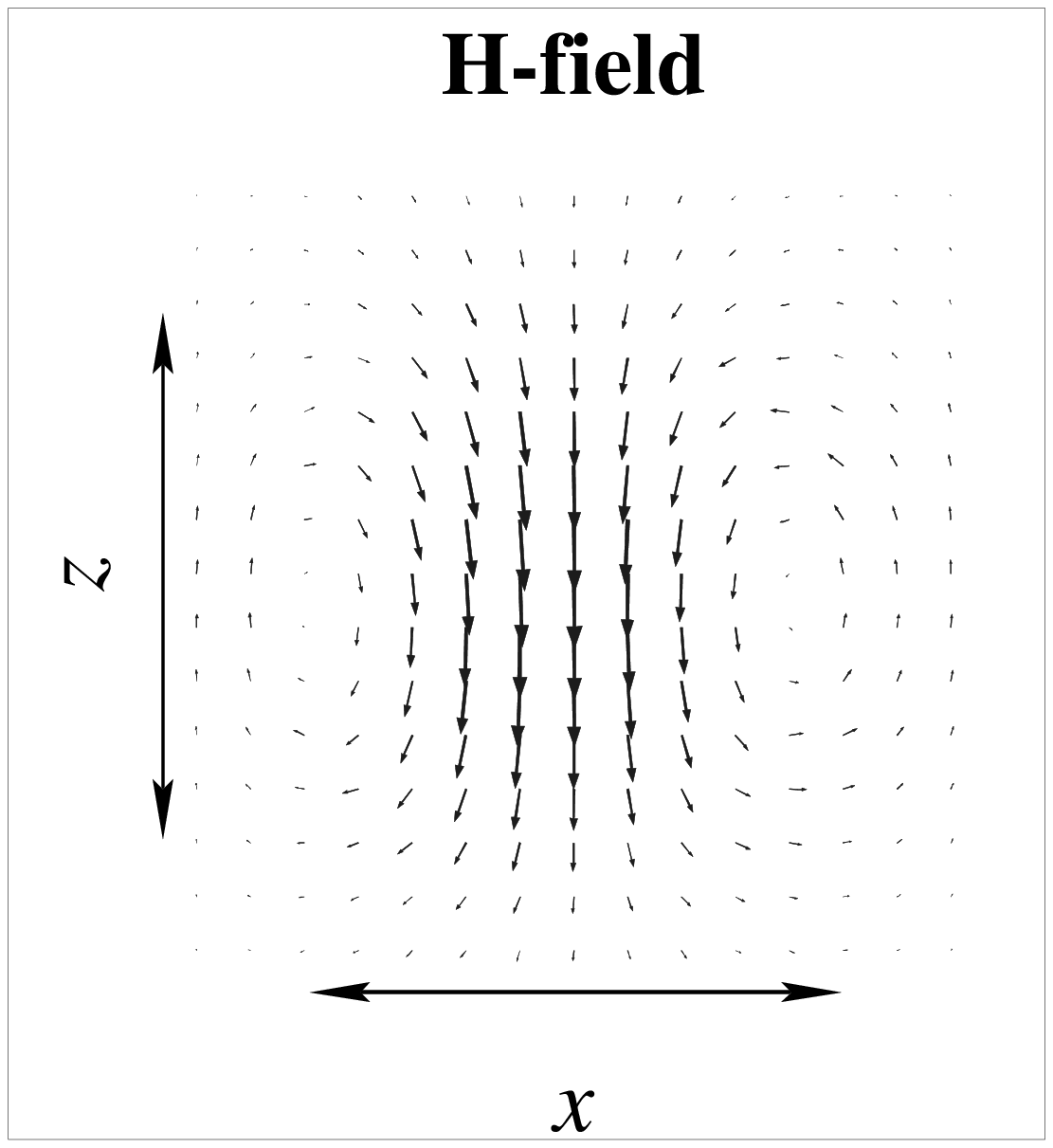}
\label{fig2a}
%\end{figure}
%
% fig2b
%
%\begin{figure}
%\special{psfile=plots/fig2/b/z8.E.s.ps hoffset=245 voffset=150 hscale=35 vscale=35}
\includegraphics{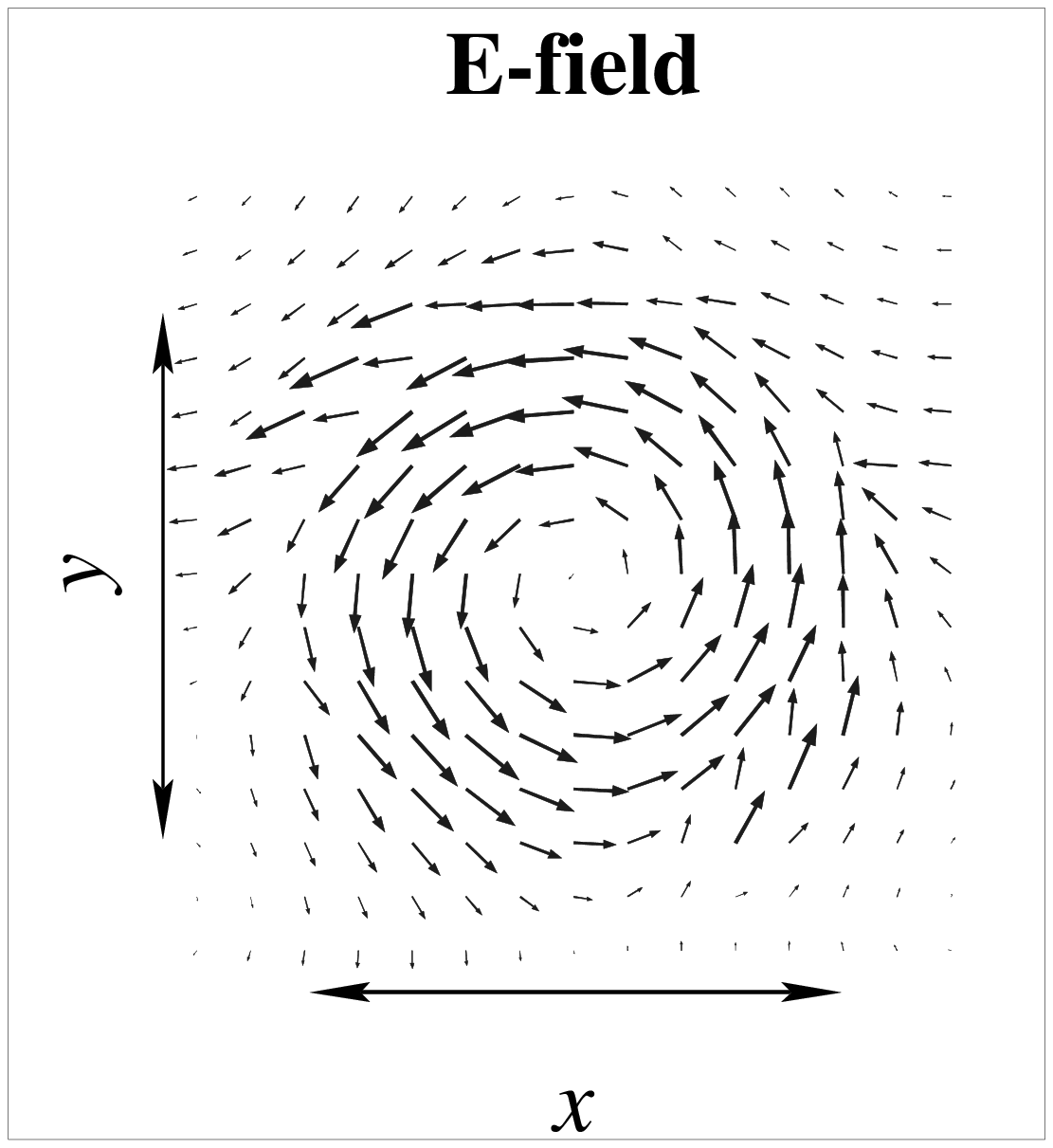}
\label{fig2b}
\caption{Field distributions within a unit cell of the photonic crystal in the lowest heavy-photon band mode. The light is $s$ polarised, normally incident from the $z$ direction. $\bf{H}$ fields are shown on a cross section $xz$ through the centre of the sphere. $\bf{E}$ fields are shown on a cross section $xy$ through the centre of the sphere. The arrows indicate the sphere edges. The field distributions are identical to those predicted by Mie theory for the $b_1^1$ mode.}
%\end{figure}
%
% fig3
%
%\begin{figure}
\epsfxsize=2in
%\epsffile{plots/fig3/p.trans.eps}
%\special{psfile=plots/fig3/p.trans.arrows.eps hoffset=190 voffset=60 hscale=70 vscale=70}
\includegraphics{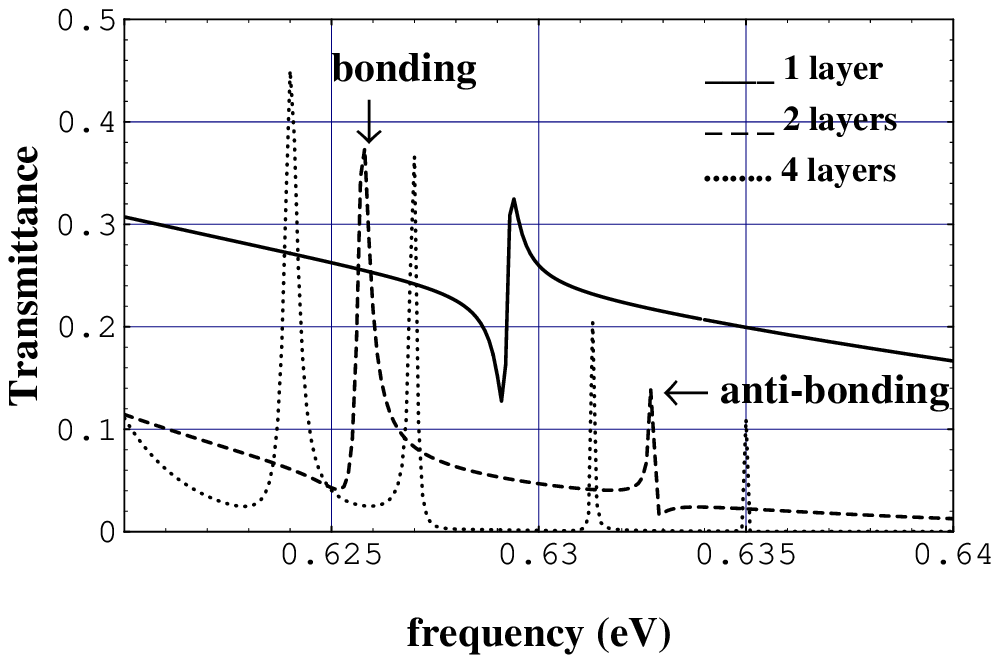}
\label{fig3}
\caption{Transmittance for three crystal samples of increasing thickness from one to four unit cells in the $z$ direction. Normal incidence.}
\end{figure}
decreases to values comparable to $l$, arising due to multiple scattering by all the lattice sites; and (c) resonant features at frequencies corresponding to the EM eigenmodes of each sphere. 

We observe these effects in {fig.1}, where the band structure (dispersion relation) is shown alongside transmittance for one layer of thickness, normal incidence. At low energies there is a proportionality between $\omega$ and the wavevector $k_z$, just as in free-wave propagation.  Forbidden ranges of $\omega$ ({\it band gaps} or {\it stop bands}) appear at higher energies. At these frequencies light is expelled from the crystal and the transmittance should fall to zero for a thick enough sample; it is impressive that even for one layer of thickness the major band gap centred at 0.7 eV causes a low transmission. The band gaps shown here along $\Gamma X$ do not extend to all other crystal directions, a fact which is typical of isolated dielectric spheres in a SC lattice \cite{sc}. Resonant effects appear as almost flat modes in the band structure combined with sharp features in transmittance. We identify such modes at frequencies 0.624,0.768,0.776,0.786,0.85,0.88 eV on the $k_z = 0$ axis. 
From now on we focus on the first resonance, appearing at $\omega_0=0.629$ eV. Being the lowest frequency resonance state it is a magnetic dipole mode. Its position on the $\omega$ axis is 20 \% too high compared to Mie theory, but this is due to the presence of nearby spheres which shift the resonance, since when we let the ratio of the lattice spacing to the sphere radius ($l/r$) rise to 10 this discrepancy decreases to a few percent. 
%
% fig4
\begin{figure}
\vskip -0.4in
%%%%%%%%\vskip -0.6in
\epsfxsize=3in
\epsfysize=3in
%\epsffile{plots/fig4/pzs.lay.eps}
%\epsffile{plots/fig4/pzs.lay.allrange.detail.eps}
\epsffile{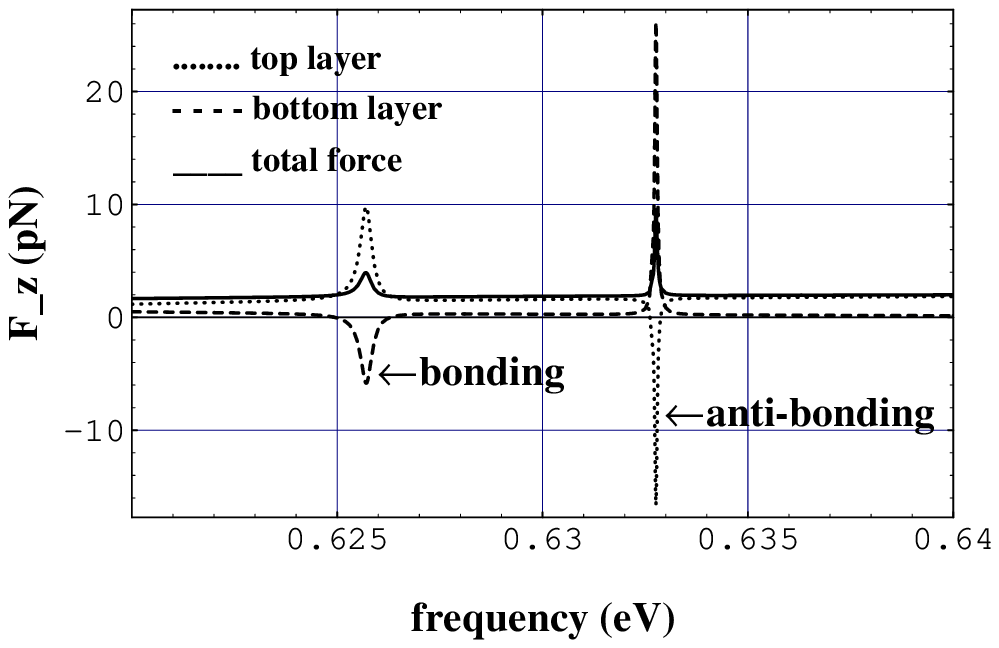}
\vskip -0.4in
%%%%%%%%\vskip -0.7in
\caption{Normal force ($F_z$) {\it per unit cell} on each layer in a two-layer system and on the whole crystal for $s$ polarisation, normal incidence. Light is incident along the $+z$ direction, so $F_z > 0 (<0)$ means positive (negative) pressure. Note that at the lower frequency forces act to push the two layers together, at the higher frequency to pull them apart.}
\label{fig4}
%\end{figure}
%
% fig5
%
%\begin{figure}
%%%%%%%%\epsfxsize=4cm
\epsfxsize=5cm
%\special{psfile=plots/fig5/Hz.bond_anti.ps hoffset=155 voffset=-10 hscale=50 vscale=50} 
\includegraphics{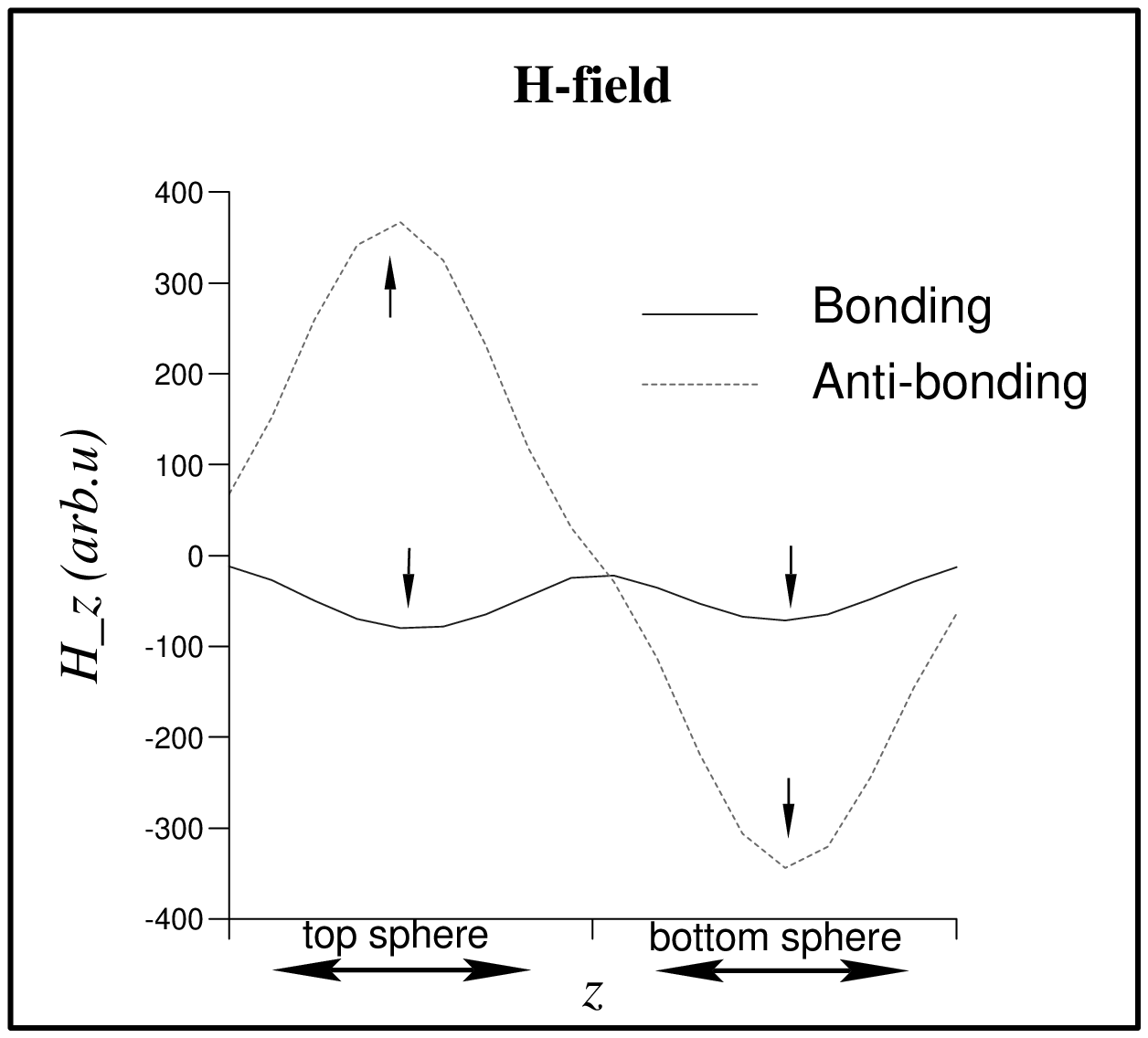} 
%%%%%%%%\special{psfile=plots/fig5/Hz.bond_anti.ps hoffset=185 voffset=10 hscale=40 vscale=40} 
\caption{Magnetic fields plotted along the $z$ axis through the centre of the spheres. Light is $s$ polarised, normally incident from the $z$ direction at frequency $\omega_<$ (bonding, dipoles parallel), $\omega_>$ (anti-bonding, dipoles anti-parallel). For $p$ polarisation the directions of the fields are reversed.}
\label{fig5}
\end{figure}
%
% fig5a
%
%\begin{figure}
%%\epsfxsize=5cm
%%\special{psfile=plots/fig5/a/bond.s.ps hoffset=110 voffset=0 hscale=35 %%vscale=40} 
%%\label{fig5a}
%\end{figure}
%
% fig5b
%
%\begin{figure}
%%\epsfxsize=5cm
%%\special{psfile=plots/fig5/b/antibond.s.ps hoffset=210 voffset=0 hscale=35 %%vscale=40} 
%%\label{fig5b}
%\end{figure}
%
% fig5c
%
%\begin{figure}
%%\special{psfile=plots/fig5/c/H.s.det.ps hoffset=5 voffset=0 hscale=30 %%vscale=30} 
%%\caption{Magnetic fields along a cross section $xz$ through the centre of %%the spheres for $s$ polarised light, normally incident from the $z$ %%direction at frequency $\omega_<$ (a), $\omega_>$ (b). For $p$ %%polarisation the directions of the fields are reversed. In (c) the fields %%in the intersphere region are shown to greater detail.}
%%\label{fig5c}
%%\end{figure}
The spatial distributions of the fields inside the spheres at $\omega_0$ are characteristic of the $b_1^1$ oscillating magnetic dipole mode of Mie theory ({figs.2a,2b}): the ${\bf H}$ field forms a hyperboloid of one sheet around which the ${\bf E}$ field circulates in a ring. So the light is exciting single sphere Mie modes. Despite the low transmittance of $\sim 0.2$, the fields are enhanced and the energy density internal to the spheres is between 2 and 3 orders of magnitude higher than that of the incident light. 

The EM forces induced on the system at the onset of resonances are quite strong. At $\omega_0$, where the transmittance displayed a sudden dip, the force peaks sharply with a Q-factor of a few thousand. 
When we increase the thickness of the sample by adding a second crystal layer, our previous expectation is verified: the sharp feature in the  transmittance and force spectra splits in two, located at frequencies slightly below ($\omega_<$) and slightly above ($\omega_>$) the original resonance ($\omega_0$). This splitting continues as we add more layers ({fig.3}): the number of peaks equals the number of layers. {Fig.4} shows the EM forces on a sample two layers thick when laser light is incident from the $z$ direction. The magnitude of the force on each layer and on the whole crystal is proportional to the intensity $I_0$ of the incident light. 
Assuming a moderate laser of $I_0 \sim 3.5 \times 10^{8}$ W/m$^2$, our calculations indicate that the photo-induced forces on each sphere range from 6 pN to 25 pN. Other forces present are Van der Waals, gravitational and thermal fluctuation forces due to Brownian motion of the spheres. The influence of thermal forces is negligible, as can be demonstrated by comparison between the thermal energy, $\frac{3}{2} k_B T \sim 6\times10^{-21}$J, and the energy of each sphere in a Mie resonance, for which a rough under-estimate is $F_{z,min} \times D \sim 6$ pN $\times170$ nm $\sim 10^{-18}$ J. Gravitational effects are also small: $m g \sim 8\times10^{-3}$ pN for GaP. 
As for the Van der Waals attraction between two GaP spheres in air ($r=365$ nm, $D=170$ nm), an upper bound for the force, calculated in the basis of pairwise additivity \cite{Israelachvili}, is $\sim 0.2$ pN. This value is clearly an overestimate, since it is obtained in the limit $D \ll r$ and ignoring retardation effects \cite{Israelachvili}\cite{Mahanty}.
Even the force between two GaP half-spaces (flat surfaces) separated by $D$ ($\sim 0.4$ pN per unit cell, according to the macroscopic Lifshitz theory  incorporating retardation and non-additivity \cite{Isr.Tabor}) is at least an order of magnitude lower than the photo-induced force. 
Why does an external source of light produce a force on each sphere so strong? The answer is that the condition for a resonance is fulfilled, greatly enhancing the internal fields. 
Due to continuity of some field components across the spherical interface,
the fields outside the sphere (within a short distance) increase also, before they start decaying rapidly. When
we integrate the stress tensor over a surface close enough to the spherical boundary, this increase yields a large force. 

The total force on the structure is positive, so the light pushes the structure as it comes along, however the forces on an individual sphere of either layer alternate sign, exhibiting a bonding/anti-bonding behaviour: at $\omega_<$ the two layers attract each other, whereas at $\omega_>$ they repel each other ({fig.4}). The magnetic fields inside each sphere indicate that at $\omega_<$ the magnetic dipoles in the top layer are parallel to the dipoles in the bottom layer, whereas at $\omega_>$ they are antiparallel ({fig.5}). Thus it is possible to interprete the forces between any two vertically adjacent spheres in the following manner. The incident EM field populates the $b_1^1$ Mie mode of each sphere, which is thus made into a magnetic dipole. Due to symmetry, the magnetic dipoles of a given layer are oriented vertically (upwards or downwards) for a normally incident beam, and they are all parallel to each other (since our system is periodic in the $xy$ plane). When a second layer is added below the first, the dipoles of the new layer have the option to orient themselves parallel or anti-parallel to the dipoles of the top layer. In the former instance there is an attraction, same in nature as the attraction between the opposite poles of two magnets (although the dipoles oscillate at the light frequency they remain always parallel and attract each other). Consequently the energy of the whole system decreases. In a similar manner when the dipoles of the bottom layer become anti-parallel to those of the top layer, repulsive interactions occur and the energy increases.

It is therefore evident that degeneracy splitting of the sharp resonance features is directly analogous to the formation of bonding and anti-bonding states of the electronic orbitals in atomic physics. Our system is the EM analogue of two Hydrogen atoms in their ground state, brought close to each other. As they approach, hybridisation occurs and the energy levels split into a bonding state at slightly lower energy ($\omega_<$, dipoles parallel), and an anti-bonding state at slightly higher energy ($\omega_>$, dipoles antiparallel) than that of the un-hybridised levels. 

This bonding/anti-bonding effect may have important implications for the fabrication of nanocrystalline materials, such as colloidal crystals or  quantum dot arrays. It may be possible to facilitate binding into 3D crystal structures, by tuning into the bonding mode, and allowing the attractive photo-induced forces to `fuse' adjacent layers together. 
When both the spheres and the reference medium have low enough absorption
(e.g. GaP spheres in water) the bonding effects are observable. Liquid media facilitate crystallisation by allowing the spheres to move slowly into an equilibrium structure; also by carrying away excess charges as well as heat which could otherwise accumulate and melt the spheres. Finally, one does not need perfect, identical spheres: objects of any shape possess their own Mie-like resonances and we have checked that the bonding effect can withstand a polydispersion of at least a few percent.

But even shining light in a solution of microspheres has a binding effect. Experiments on gold nanospheres by Kimura \cite{Kim} have shown fast formation of aggregates due to photo-enhanced Van Der Waals forces. Burns {\it et al} have experimentally demonstrated the presence of attractive forces, induced by laser light, between two dielectric spheres in water \cite{Golov}. 
They argue that these are magnetic dipole interactions, owing their existence to induced currents on the spheres, and dominating all other forces when the vector separation between the oscillating currents is perpendicular to the electric field polarisation, in exact agreement with our results.

What happens for the Mie modes at higher frequencies? Whenever the frequency of the incident light lies within a heavy-photon band which is of Mie origin, resonance phenomena are expected. The transmittance for a crystal one layer thick will therefore display sharp peaks (dips), which for thicker crystals will be enhanced (reduced), when the band is degenerate (non-degenerate) \cite{SKM}. The field distributions will be characteristic of the particular mode excited which will in general be of electric or magnetic type. Sharp features will also arise in the forces between spheres. This situation is expected to be especially relevant for the few lowest heavy-photon modes, for the reason that the density of Mie resonances per frequency is lowest then. As $\omega$ increases, there are more different Mie modes lying close together. Because each Mie-originating band has a finite dispersion, the presence of many of these bands in neighbouring frequency regions is expected to lead into hybridisation so that many spherical waves will be mixing within the same heavy-photon band. When this happens, the sharpness of resonance effects will be reduced. 

In conclusion, we have studied the influence of the Mie-originating heavy-photon modes on forces in nanostructures. In a lattice of dielectric spheres, the lowest in frequency such mode is an oscillating magnetic dipole. When EM radiation at the frequency of this mode is incident on the crystal, resonance effects enhance the fields internal to the spheres, resulting in intense forces between the crystal layers. Two spheres of adjacent layers respond to radiation in a way which is strikingly similar to the atomic physics bonding/anti-bonding idea of how two identical atoms interact when coming close together. Depending on the laser power and the dielectric properties of the spheres, this photo-induced force can exceed gravitational, Van der Waals and other forces present and play a dominant role in cohesion in nanostructures. Apart from possible applications in the fabrication of 3D photonic crystals, the resonant mechanism leading to these forces may contribute to our understanding of novel non-linear phenomena arising due to the application of laser light fields in nanostructures. 

%
%%%   In the Acknoledgments, use the following macro  before and instead 
%%%   of  ``Acknoledgments''
%
\stars
M.I.A. gratefully acknowledges stimulating discussions with Francisco-Jos{\'e} Garcia-Vidal.
%
%%%   Bibliography environment begins here. You can use the macros \Name{},
%%%   \And, \Book{} or \Review{}, \Vol{}, \Year{} and \Page{}, to type your
%%%   references.
%

\end{document}